\documentstyle[sprocl,epsf]{article} 
\def\dK{\mathop{\rm K}}
\def\etal{{\em et al}\/}
\def\max{\mathop{\rm max}}
\def\N{\mathop{\rm N}}
\def\pr{\mathop{\rm p}}
\def\stat{\mathop{\rm stat}}
\def\syst{\mathop{\rm syst}}

\def\lesssim{\mathbin{\;\raise1pt\hbox{$<$}\kern-8pt\lower3pt\hbox{$\sim$}\;}}
\def\gtrsim{\mathbin{\;\raise1pt\hbox{$>$}\kern-8pt\lower3pt\hbox{$\sim$}\;}}
\def\95cl{\mathop{95 \%\ \rm c.l.}}

\def\H{\mathop{\rm H}}
\def\Htwo{\mathop{\rm D}}
\def\Hethree{^3{\rm He}}
\def\Hefour{^4{\rm He}}
\def\Lisix{^6{\rm Li}}
\def\Liseven{^7{\rm Li}}

\begin{document}

\title{PRIMORDIAL NUCLEOSYNTHESIS AND DARK MATTER}

\author{\sc Subir Sarkar}

\address{Theoretical Physics, University of Oxford, 
         1 Keble Road, Oxford OX1 3NP, U.K.}

\maketitle\abstracts{The cosmological abundance of nucleons determined
from considerations of big bang nucleosynthesis allegedly provides
compelling evidence for non-nucleonic dark matter. Recent developments
in measurements of primordial light element abundances, in particular
deuterium and helium, require reexamination of this important
issue. The present situation is uncertain but exciting.}

\small
\noindent
``At this point I would like to make a remark on the present state of
observations relevant to cosmology. When a physicist reads a paper by
a typical astronomer, he finds an unfamiliar style in the treatment of
uncertainties and errors \ldots The authors are apparently unwilling
to state precisely the odds that their number is correct, although
they have pointed out very carefully the many sources of error, and
although it is quite clear that the error is a considerable fraction
of the number. The evil is that often other cosmologists or
astrophysicists take this number without regard to the possible error,
treating it as an astronomical observation as accurate as the period
of a planet.''
\vspace{-2mm}
\begin{flushright}
Richard Feynmann (1962)\cite{feyn}
\end{flushright}
\normalsize
\vspace{-5mm}

\section{Introduction}
	
The focus of this Workshop is the problem of dark matter. It is well
established that the dynamics of galaxies and clusters is dominated by
unseen matter which contributes $\sim10-20\%$ of the critical
density.\cite{book} In contrast, the luminous (nucleonic) matter in
such structures has a density parameter of only \cite{persal}
\begin{equation}
\label{omeganps}
 \Omega_{{\N}} \simeq 2.2 \times 10^{-3} + 6.1 \times 10^{-4} h^{-1.3}\ ,
\end{equation}
where the first term accounts for the stars and the second for the
X-ray emitting gas. Here,
$h\equiv\,H_0$/100\,km\,sec$^{-1}$\,Mpc$^{-1}$ is the present Hubble
parameter. The obvious question is whether the dark matter might also
be nucleonic but in the form of say cold compact objects or diffuse
gas.\cite{carr}

We also recognize a second, indirect, dark matter problem in that the
standard Friedmann-Robertson-Walker cosmology must be incredibly fine
tuned if the universe does not have exactly the critical density, as
would be naturally ensured by an initial De Sitter (inflationary)
epoch.\footnote{In principle inflation would be consistent with a
present day cosmological constant, requiring only that $\Omega_{\rm
matter}+\Omega_{\Lambda}=1$. However for the two contributions to be
comparable today would also require severe fine tuning, hence for
consistency we must assume $\Omega_{\Lambda}=0$.} Then there must be a
substantial amount of matter in some form which is not clustered on
the scales probed by dynamical measurements of galaxies or
clusters. Indeed studies \cite{dekel} of non-Hubble velocity flows on
larger scales indicate that $\Omega>0.3$ and recent attempts
\cite{perl} to measure the large-scale curvature of space-time using
Type\,I SN as `standard candles' find $\Omega>0.5$. Finally,
degree-scale measurements of the cosmic microwave background (CMB)
anisotropy \cite{cmb} provide preliminary evidence of a `Doppler peak'
at the position expected for a critical density universe.

Recently there has been considerable progress on the first dark matter
problem through searches \cite{macho} for microlensing which indicate
\cite{wyn} that $\approx40\%$ of the dark matter in the halo of our
Galaxy is in the form of compact, presumably nucleonic, objects of
mass $\sim0.1-1$\,M$_\odot$. Given the plethora of exotic particle
candidates for the dark matter in extensions of physics beyond the
Standard Model \cite{robi} and the increasing number of experimental
efforts at their direct detection, it is clearly crucial
to establish whether there really is any hard evidence for
non-nucleonic dark matter. It is here that considerations of Big Bang
nucleosynthesis (BBN) play a crucial role since the abundances of the
light elements provide a limit on the abundance of nucleons
\footnote{We distinguish between nucleons and baryons since there may
well be baryonic dark matter (e.g. strange quark nuggets, black holes)
which does not participate in BBN.} in {\em any} form. This limit on
$\Omega_{\N}$ has also become relevant to the indirect dark matter
problem because of {\sl ROSAT} X-ray observations which reveal rather
large nucleon fractions in clusters of galaxies. For example in the
{\it Coma} cluster,\cite{white}
\begin{equation}
\label{fn}
 f_{\N} \equiv \frac{M_{\N}}{M_{\rm tot}} \geq 0.009 + 0.05\,h^{-3/2}\ ,
\end{equation}
where the first term accounts for the stellar matter within the Abell
radius of $r_{\rm A}\approx1.5\,h^{-1}$ Mpc and the second for the
X-ray emitting gas. Hydrodynamical simulations of cluster formation in
a CDM universe find that cooling and other dissipative effects could
have enhanced $f_{\N}$ within $r_{\rm A}$ by a factor of at most
$\Upsilon\approx1.4$ over the global average.\cite{white} Since the
global density parameter is just
$\Omega=\Upsilon(\Omega_{\N}/f_{\N})$, we see that $\Omega=1$ would be
permitted only if $\Omega_{\N}\gtrsim0.1$.

Although these are well known issues they have become somewhat
controversial of late since the quoted limit on $\Omega_{\N}$ has
varied depending on how different authors have inferred the primordial
elemental abundances from their present values, using models of
galactic chemical evolution. In fact some authors \cite{hata} have
gone so far as to question the consistency of standard BBN itself. We
have argued elsewhere \cite{kersar} that there is {\em no} such crisis
if a conservative view is taken of observational errors. Physicists
are often puzzled that such very different conclusions can be drawn on
the basis of (presumably) the same observational data. The words of
Feynmann \cite{feyn} quoted above appear to be as relevant today as
when they were spoken 34 years ago!

\section{Big Bang Nucleosynthesis}

\begin{figure}[tbhp]
\epsfxsize\hsize\epsffile{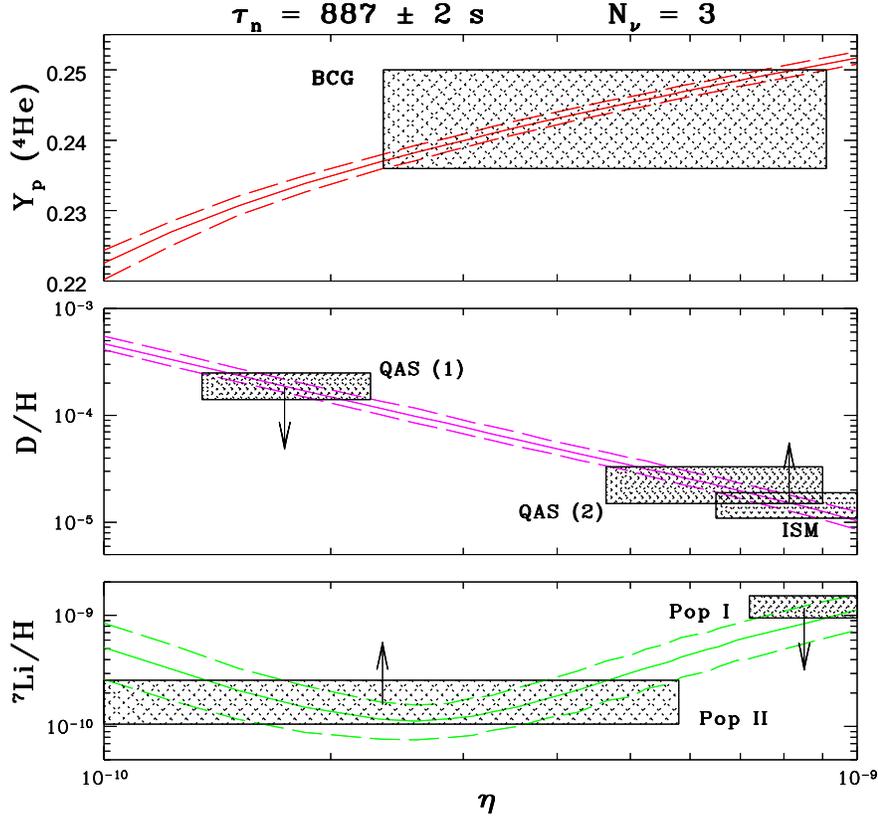}
\caption{Predicted light element abundances for the Standard Model
 (with $N_\nu=3$ neutrino species) versus the nucleon-to-photon ratio
 $\eta$.$^{13}$ The $\95cl$ limits determined by Monte Carlo reflect
 the uncertainties in all input nuclear cross-sections and the neutron
 lifetime. Rectangles indicate the various observational
 determinations and associated `$\95cl$' bounds. The $\Hefour$
 abundance is obtained from observations of metal-poor blue compact
 galaxies by linear extrapolation to zero metallicity;$^{28}$ the
 upper bound is reliable, the lower one less so. Both the $\Htwo$
 abundance measurements $^{44,50}$ in quasar absorption systems are
 shown; the higher value is interpreted as an upper bound. Also shown
 is the abundance in the interstellar medium $^{31}$ which provides a
 reliable lower bound. The $\Liseven$ abundance as measured in the
 hottest, most metal-poor halo stars $^{55}$, as well as in disk stars
 $^{58}$ is shown and interpreted as providing, respectively, reliable
 lower and upper bounds on its primordial value. GIven these
 uncertainties, the Standard Model is presently consistent with
 observations for $\eta$ in the range $\sim(2-9)\times10^{-10}$.}
\label{abund}
\end{figure}

The physics of BBN is well understood \cite{bbn} and in recent years
the uncertainties in the input nuclear reaction cross-sections and the
neutron lifetime have been included in the computer code using Monte
Carlo methods, thus accounting for all correlated
effects.\cite{smith,kraker} As shown in figure~\ref{abund}, the
$\Hefour$ abundance is known to within $\pm0.5\%$,\footnote{However
small corrections to the $\Hefour$ abundance for finite temperature
effects \cite{myrev} may not have quite stabilized yet \cite{sawyer}
so there {\em could} be a comparable systematic uncertainty.} but the
$\Htwo$ and $\Hethree$ abundances have a $\pm15\%$ uncertainty, while
the $\Liseven$ abundance is uncertain to within $\pm60\%$. There are
of course many possibilities for departures from the standard model,
e.g. an inhomogeneous nucleon distribution or non-zero neutrino
chemical potentials.\cite{exotic} However recent developments in our
understanding of cosmological phase transitions and lepto/baryogenesis
\cite{shaposh} do not motivate such non-standard scenarios and
moreover they are highly constrained by the observational data. It is
therefore reasonable to adopt the standard picture which has only two
unknown parameters, viz. the nucleon-to-photon ratio,
$\eta\equiv\,n_{\N}/n_{\gamma}\simeq2.72\times10^{-8}\Omega_{\N}h^2(T_0/2.73\dK)^{-3}$,
and the effective number of massless, 2-component neutrinos,
$N_\nu$.\footnote{The number of left-handed doublet neutrinos is known
to be 3 from {\sl LEP}, but we must allow for the possibility that
there are other relativistic particles (e.g. singlet neutrinos) which
may contribute to the energy density, and parametrize their
contribution in terms of an effective $N_\nu$. We do not entertain the
possibility that the $\nu_\tau$ is unstable and decays before BBN as
this is ruled out experimentally if the decays are into known
particles.\cite{myrev}} The latter determines the expansion rate of
the universe (hence the free neutron abundance when nucleosynthesis
begins) while the former determines the rates of nuclear reactions
(which synthesize essentially all neutrons into $\Hefour$ nuclei,
leaving behind small traces of $\Htwo$, $\Hethree$ and $\Liseven$).
Therefore the $\Hefour$ abundance increases proportionally as $N_\nu$ but
only logarithmically as $\eta$ while the other elemental abundances
are strongly dependent on $\eta$, as seen in figure~\ref{abund}.

The essential problem in attempting to compare the theoretical
predictions with observational data is that the primordial abundances
have been significantly altered during the lifetime of the universe
through nuclear processing in stars and other galactic chemical
evolution effects.\cite{rev} The most stable nucleus, $\Hefour$, grows
in abundance with time since it is always created in stars, while
$\Htwo$, the most weakly bound, is always destroyed. The history of
$\Hethree$ and $\Liseven$ is more complicated since these elements may
be both destroyed and created. To avoid uncertain corrections, it is
neccessary to measure abundances in the most primordial material
available and the recent development of large telescopes and CCD
imaging technology have led to significant progress in the field. We
present the key results and recent developments below and refer for
more details to a recent review.\cite{myrev}

\subsection{Helium-4}

\begin{figure}[tbh]
\mbox{\epsfxsize6cm\epsffile{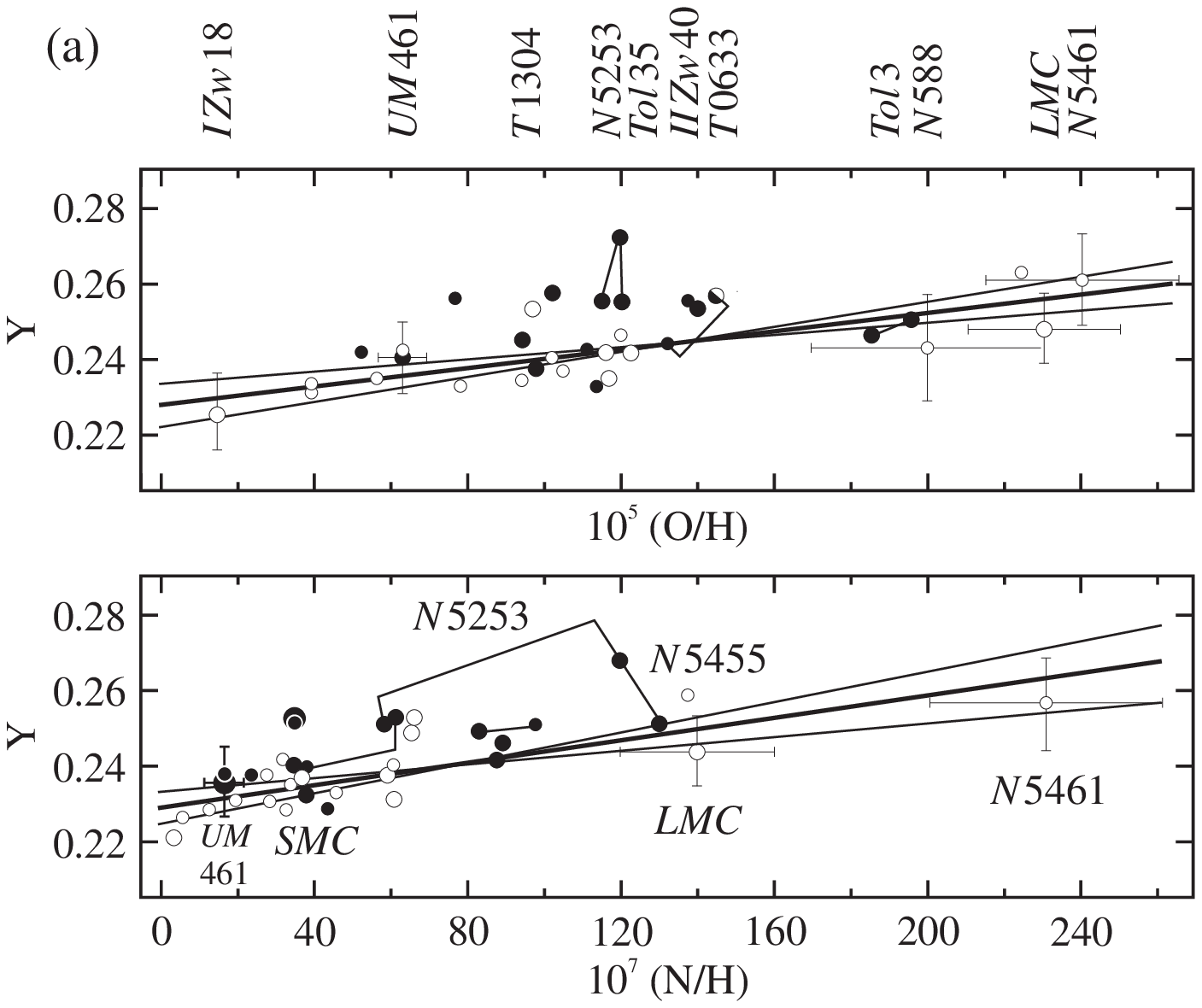}
      \epsfxsize6cm\epsffile{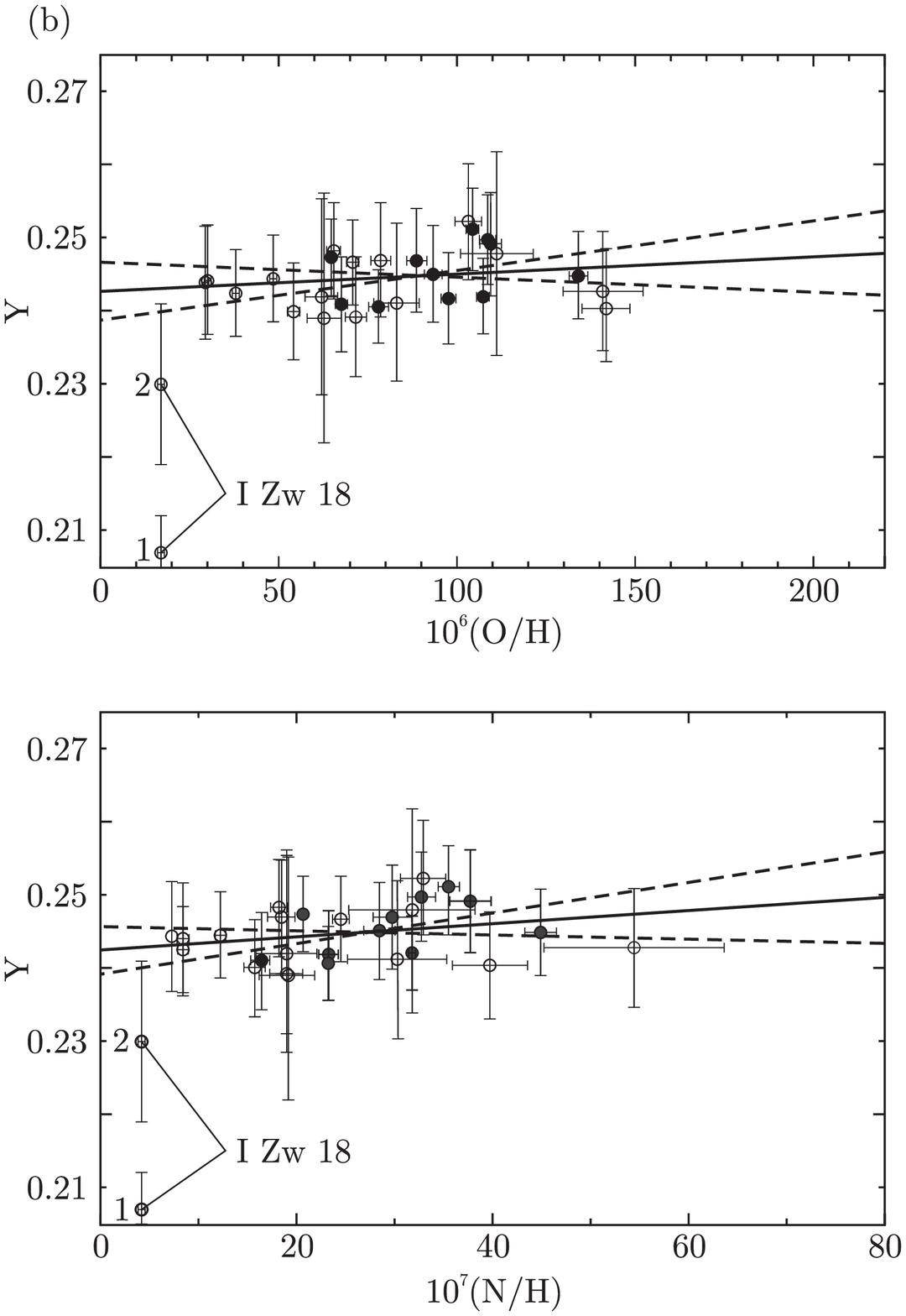}}
\caption{Regressions of the helium mass fraction against the oxygen
 and nitrogen abundances in extragalactic low-metallicity H\,II
 regions, with (filled circles) and without (open circles) broad
 Wolf-Rayet features. Panel (a) shows abundances for 33 objects
 obtained using the Brocklehurst emissivities,$^{22}$ with the
 maximum-likelihood linear fits (with $\pm1\sigma$ limits) for the
 latter category. Panel (b) shows abundances for 27 objects obtained
 using the Smits emissivities,$^{28}$ along with the
 maximum-likelihood linear fits (with $\pm1\sigma$ limits).}
\label{Ypreg}
\end{figure}

To determine the primordial $\Hefour$ mass fraction, $Y_{\pr}$, we
must allow for stellar helium production through its correlation with
elements which are made only in stars. This is best done by studying
recombination lines from H\,II regions in blue compact galaxies (BCGs)
where relatively little stellar activity has occured, as evidenced by
their low `metal' abundance. The data set of Pagel \etal~\cite{pagel}
gathered from 33 selected objects indicated a primordial abundance of
$Y_{\pr}({\Hefour})=0.228\pm0.005$ (with an estimated systematic error
of $\pm0.005$.) As shown in figure~\ref{Ypreg}(a) this is obtained by
linear extrapolation to zero metal abundance in a plot of the measured
helium abundance against that of oxygen and nitrogen. A similar result
was obtained for another set of 11 BCGs by Skillman \etal.\cite{skill}
Olive and Steigman \cite{os} made a fit to a selected subset of the
combined data and quoted
\begin{equation}
\label{Ypos}
 Y_{\pr}({\Hefour}) = 0.232 \pm 0.003\ (\stat) \pm 0.005\ (\syst)\ .
\end{equation} 

However, it has been argued that the systematic error may be
significantly larger than the value estimated above.\cite{syst} A
particular issue is the input atomic physics used to extract the
abundance from the measured line strengths. The point is that the
He\,I line intensities deviate from the pure recombination values, due
mainly to collisional excitation from the metastable 2$^{3}S$
state. To correct for this, the physical conditions, in particular the
electron density, in the H\,II regions must be accurately known. This
is best done by simultaneously measuring several lines and determining
the corrections self-consistently by demanding that all line ratios
have their recombination values after correction. Izotov
\etal~\cite{izotov} have recently done this in a study of 27 objects,
using data on 4 different lines (including the triplet line
$\lambda7065$). They used the updated emissivities of Smits
\cite{smits} which are $\approx50\%$ higher for this line than those
of Brocklehurst \cite{brock}, which were used in previous work. As
shown in figure~\ref{Ypreg}(b) they find a higher intercept of
$Y_{\pr}({\Hefour})=0.243\pm0.003$, with a smaller dispersion of the
data points in their regression plots. Moreover the derived slope,
${\rm d}Y/{\rm d}Z\approx1.7\pm0.9$, is smaller than the value of
$6.1\pm2.1$ found earlier,\cite{pagel} in better agreement with
general theoretical expectations. Izotov \etal\ argue that other
possible systematic effects (in particular corrections for fluorescent
enhancement in the $\lambda7065$ line) cannot alter $Y_{\pr}$ by more
than $\pm0.001$ so that
\begin{equation} 
\label{Ypitl}
  Y_{\pr}({\Hefour}) = 0.243 \pm 0.003\ (\stat) \pm 0.001\ (\syst)\ .
\end{equation}
This value is presently controversial; for example in his talk
\cite{keith} at this meeting, Keith Olive criticized their use of the
$\lambda7065$ line in view of its sensitivity to collisional
excitation. However if it is excluded the correction for collisional
enhancement would depend, as in previous work, on more uncertain
estimates of the electron density from nebular S~II emission lines! I
recall the admonishment of Zeldovich that ``Observations should be
analyzed by astronomers and not by theoretical
physicists''.\cite{zeld} However I agree with Olive that the Izotov
\etal\ result is higher because of dropping the lowest metallicity
galaxy {\it I\,Zw\,18} from their sample (see
figure~\ref{Ypreg}(b)). They did so on account of its anomalously low
He\,I line intensities, said to be caused by underlying stellar and
interstellar absorption; whether this is justified or not is for other
{\em observers} to decide. I maintain that while more work is
neccessary to settle this issue, it would be {\em conservative} at
this time to allow for the possibility that the upper bound to
$Y_{\pr}({\Hefour})$ is 0.25 rather than 0.24. The lower bound is less
certain since the {\em linear} extrapolation to zero metallicity is
purely empirical.

\subsection{Deuterium and Helium-3}

Deuterium is detected in the local interstellar medium through its
ultraviolet absorption lines in stellar spectra but, as expected for a
fragile element, its abundance shows a large scatter,
${\Htwo}/{\H}\approx(0.2-4)\times10^{-5}$, suggesting localized
abundance fluctuations and/or systematic errors. McCullough
\cite{mccull} finds that after discarding some unreliable
measurements, the 7 {\sl IUE} and 14 {\sl Copernicus} measurements along
the cleanest lines of sight (towards hot stars within about 1 kpc) are
all consistent with an interstellar abundance of
\begin{equation}
\label{Dism}
 \left(\frac{\Htwo}{\H}\right)_{\rm ISM} = 1.5 \pm 0.2 \times 10^{-5}\ .
\end{equation}
Recently the {\sl HST} has provided a more accurate measurement of
${\Htwo}/{\H}=1.60\pm0.09\,(\stat)
^{+0.05}_{-0.10}\,(\syst)\times10^{-5}$ towards the star {\it Capella}
at 12.5~kpc.\cite{linsky} However since the Lyman-$\alpha$ line (of
hydrogen) is severely saturated even towards such a nearby star, such
observations, although precise, cannot test whether there are real
spatial variations in the interstellar deuterium abundance. It has
been argued \cite{epstein} that there are no important astrophysical
sources of deuterium and observational attempts to detect signs of
deuterium synthesis in the Galaxy have so far not contradicted this
belief.\cite{pas} Then the lowest $\Htwo$ abundance observed today
provides a reliable lower bound to the primordial abundance,
viz. $({\Htwo}/{\H})_{\pr}>1.1\times10^{-5}$.

There are similar large fluctuations in the abundance of $\Hethree$
which has been detected through its radio recombination line in a
dozen galactic H\,II regions. The values measured by Balser \etal
\cite{balser} range over
\begin{equation}
\label{He3ism}
 \left(\frac{\Hethree}{\H}\right)_{\rm H\,II} \sim (1-4) \times 10^{-5}\ .
\end{equation}
It has also been detected \cite{rood} with a large abundance
(${\Hethree}/{\H}\approx10^{-3}$) in the planetary nebula {\it
NGC3242}, in accord with the theoretical expectation \cite{dear} that
it is created in low mass stars. However the galactic observations
find the highest $\Hethree$ abundances in the outer Galaxy where
stellar activity is {\em less} than in the inner Galaxy. While regions
with high abundances do lie preferentially in the Perseus spiral arm,
there are large source-to-source variations which do not correlate
with stellar activity.\cite{balser} Thus these measurements do not
provide any reliable cosmological input.

Yang \etal~\cite{yang} had suggested that the uncertainties in
determining the primordial abundances of $\Htwo$ and $\Hethree$ may be
circumvented by considering their {\em sum}. They argued that since
$\Htwo$ is burnt in stars to $\Hethree$, a fraction $g_{3}$ of which
survives stellar processing, the primordial abundances may be related
to the abundances later in time through the inequality
\begin{equation}
\label{yangevol}
  \left(\frac{{\Htwo} + {\Hethree}}{\H}\right)_{\pr} < 
   \left(\frac{{\Htwo} + {\Hethree}}{\H}\right) +
   \left(\frac{1}{g_{3}} - 1\right) \left(\frac{\Hethree}{\H}\right)\ .
\end{equation}
As reviewed by Geiss,\cite{geiss} the terms on the rhs may be
determined at the time of formation of the Solar system, 4.6\,Gyr ago.
The abundance of $\Hethree$ in the Solar wind, deduced from studies of
gas-rich meteorites, lunar rocks and metal foils exposed on lunar
missions, may be identified with the sum of the pre-Solar abundances
of $\Hethree$ and $\Htwo$ (which was burnt to $\Hethree$ in the Sun),
while the smallest $\Hethree$ abundance found in carbonaceous
chondrites, which are believed to reflect the composition of the
pre-Solar nebula, may be identified with the pre-Solar abundance of
$\Hethree$ alone. For example, Walker \etal~\cite{walker} obtained
\begin{equation}
\label{Solarsys}
 1.3 \times 10^{-5} \lesssim \left(\frac{\Hethree}{\H}\right)_{\odot} 
     \lesssim 1.8 \times 10^{-5}\ , \quad
 3.3 \times 10^{-5} \lesssim \left(\frac{{\Htwo}+{\Hethree}}{\H}\right)_{\odot}
  \lesssim 4.9 \times 10^{-5} .
\end{equation}
These authors also interpreted a study \cite{dear} on the survival of
$\Hethree$ in stars to imply the lower limit $g_{3}\gtrsim0.25$.
Using these values yields the bound \cite{yang,walker}
\begin{equation}
\label{He3plusD}
 \left(\frac{{\Htwo}+{\Hethree}}{\H}\right)_{\pr} \lesssim 10^{-4}\ ,
\end{equation}
which is essentially a bound on primordial $\Htwo$ alone since it is
relevant only at small $\eta$ where the relative abundance of
$\Hethree$ is negligible.

There are however several reasons to distrust the above bound, from
which a stringent lower limit on $\eta$ has been
deduced.\cite{yang,walker,smith,copi1} First, it is not clear if the
Solar system abundances provide a representative measure at all, given
that observations of $\Hethree$ elsewhere in the Galaxy reveal
unexplained source-to-source variations. Indeed the pre-Solar
abundance of $\Hethree$ is {\em less} than some of the present day
interstellar values. Second, the survival fraction of $\Hethree$ may
have been overestimated since there may be net {\em destruction} of
$\Hethree$ in low mass stars through the same mixing process which
appears to be needed to explain other observations, e.g. the
$^{12}$C/$^{13}$C ratio.\cite{hogan} In fact a recent measurement
using {\sl Ulysses} finds that
${\Hethree}/{\Hefour}=2.2^{+0.7}_{-0.6}\,(\stat)\,\pm0.2\,(\syst)\times
10^{-4}$ in the local interstellar cloud, rather close to the value of
$1.5\pm0.3\times10^{-4}$ in the pre-solar nebula, demonstrating that
the $\Hethree$ abundance has hardly increased since the formation of
the Solar system.\cite{glogei}

It is obviously crucial to detect deuterium outside the Solar system
and the nearby interstellar medium in order to get at its primordial
abundance and also, of course, to establish its cosmological origin.
Astronomers have attempted to measure Lyman-series absorption lines of
deuterium in the spectra of distant quasars, due to foreground
intergalactic clouds made of primordial unprocessed material. Problems
arise in studying such quasar absorption systems (QAS) because of
possible confusion with neighbouring absorption lines of hydrogen and
multi-component velocity structure in the clouds. The advent of large
aperture ground-based telescopes, e.g. the 10-mt {\sl Keck Telescope},
has provided the required sensitivity and spectral resolution, leading
to several detections. Songaila \etal~\cite{songaila} find
\begin{equation}
\label{Dlya1}
 \left(\frac{\Htwo}{\H}\right)_{\rm QAS (1)} \approx (1.9-2.5)\times 10^{-4}\ ,
\end{equation}
in a chemically unevolved cloud at $z=3.32$ along the line of sight to
the quasar {\it Q0014+813}, and note that there is a $3\%$ probability
of the absorption feature being a misidentified Ly-$\alpha$ line of
hydrogen. Further observations by Rugers and Hogan \cite{rughoga} have
resolved $\Htwo$ lines at $z=3.320482$ and $z=3.320790$, thus
eliminating the possibility of such confusion; the measured abundances
in the two clouds are, respectively, ${\Htwo}/{\H}=10^{-3.73\pm0.12}$
and $10^{-3.72\pm0.09}$ (where the errors are {\em not} gaussian). An
independent lower limit of ${\Htwo}/{\H}\geq1.3\times10^{-4} (\95cl)$
is also set on their sum from the Lyman limit opacity. Recently, these
authors have detected ${\Htwo}/{\H}=1.9^{+0.6}_{-0.9}\times10^{-4}$ in
another QAS at $z=2.797957$ towards the same quasar; The errors are
higher because the ${\Htwo}$ feature is saturated, even so a $\95cl$
lower limit of ${\Htwo}/{\H}>0.7\times10^{-4}$ is
obtained.\cite{rughogb} There have been other, less definitive,
observations of QAS consistent with this abundance,
e.g. ${\Htwo}/{\H}\lesssim1.5\times10^{-4}$ at $z=4.672$ towards {\it
BR1202-0725},\cite{wampler} and ${\Htwo}/{\H}\lesssim10^{-3.9\pm0.4}$
at $z=3.08$ towards {\it Q0420-388}.\cite{carswell} However, very
recently, Tytler and collaborators have found much lower values in QAS
at $z=3.572$ towards {\it Q1937-1009} \cite{tytler} and at $z=2.504$
towards {\it Q1009+2956};\cite{burles} their average abundance is
\begin{equation}
\label{Dlya2}
 \left(\frac{\Htwo}{\H}\right)_{\rm QAS (2)} = 2.4 \pm 0.3\ (\stat) \pm 0.3\ 
                                               (\syst) \times 10^{-5}\ .
\end{equation}
Since the Lyman-$\alpha$ line (of hydrogen) is saturated in these
objects, the $\Htwo$ abundance must be derived from a cloud model,
with associated uncertainties.\cite{crit} A recent independent
measurement \cite{cowie} of the H\,I density in the cloud towards {\it
Q1937-1009} raises the deuterium abundance to
${\Htwo}/{\H}\gtrsim4\times10^{-5}$. The situation is clearly volatile
at the moment! We adopt the {\em conservative} viewpoint that the
`ceiling' to the ${\Htwo}$ measurements in QAS provides a reliable
upper bound to its primordial abundance,
viz. $({\Htwo}/{\H})_{\pr}<2.5\times10^{-4}$.

\subsection{Lithium-7}

Lithium is observed in both halo (Pop\,II) and disk (Pop\,I) stars,
with widely differing abundances.\cite{lirev} For Pop\,I stars in open
clusters with ages up to 10\,Gyr, the observed abundances range upto
${\Liseven}/{\H}\sim10^{-9}$. However in the older Pop\,II halo
dwarfs, Spite and Spite \cite{spite} observed the abundance to be
$\approx10$ times lower and, for high temperatures and low
metallicity, nearly independent of the stellar temperature and the
metal abundance. This has been used to argue that the Pop\,II
abundance reflects the primordial value in the gas from which the
stars formed, with the higher abundance in the younger Pop\,I stars
created subsequently. For example Walker \etal~\cite{walker} took the
primordial abundance to be an weighted average of the data for 35
stars, $({\Liseven}/{\H})_{\pr}=10^{-9.92\pm0.07}(\95cl)$. However
Thorburn \cite{thor} finds a weak trend of increasing $\Liseven$
abundance with both increasing temperature and increasing metallicity
in a larger sample of 80 stars. The elements beryllium and boron are
also observed in similar stars, correlated with the metallicity and in
the ratio B/Be\,$\approx10$, which indicates that they were produced
by galactic cosmic ray spallation rather than being
primordial.\cite{gilmore} This should have also made $\approx35\%$ of
the observed $\Liseven$. The primordial abundance can then be
identified \cite{thor} with the average value in the hottest, most
metal-poor stars, viz.
\begin{equation}
\label{Li7popII}
  \left(\frac{\Liseven}{\H}\right)_{\pr}^{\rm II} = 10^{-9.78 \pm 0.20} 
   \ (\95cl)\ .
\end{equation}
However Molaro \etal~\cite{molaro} do not find any correlation with
the temperature or metallicity in another sample of 24 halo dwarfs
using different modelling of the stellar atmospheres. The abundance
they obtain is fortuitously identical to that given above but this
does highlight the systematic uncertainties involved. Moreover there
are several Pop\,II halo dwarfs which have {\em no} detectable
lithium.\cite{thor} This suggests that the primordial $\Liseven$
abundance may instead be the Pop\,I value which has been depleted down
to (and occasionally, below) the level in Pop\,II stars, for example
through turbulent mixing driven by stellar rotation. Stellar modelling
shows that the primordial abundance can then be as high as
\cite{chabdem}
\begin{equation}
\label{Li7popI}
  \left(\frac{\Liseven}{\H}\right)_{\pr}^{\rm I} = 10^{-8.92 \pm 0.1} .
\end{equation}
An argument against such severe depletion is that $\Lisix$, an even
more fragile isotope, has been detected \cite{smith2} in one of the
hottest known Pop\,II stars with $({\Lisix}/{\Liseven})^{\rm
II}=0.05\pm0.02$. However it is possible, e.g. through mass loss by
stellar winds, for $\Liseven$ to be depleted without depleting
$\Lisix$; the preferred primordial abundance would then be the upper
envelope of the Pop\,II value.\cite{vauchar}

\section{Theory versus observations}

In figure~\ref{abund} we show that the standard model with $N_\nu = 3$
is consistent with these observations over a wide range of
$\eta\sim(2-9)\times10^{-10}$. The value of $\eta$ will be close to
its minimum allowed value if the high $\Htwo$ abundance in QAS
(eq.\ref{Dlya1}) and the Pop\,II $\Liseven$ abundance
(eq.\ref{Li7popII}) are primordial, while it will be close to its
maximum allowed value if instead the low $\Htwo$ abundance in QAS
(eq.\ref{Dlya2}) and the Pop\,I $\Liseven$ abundance
(eq.\ref{Li7popI}) are primordial. Of course a value of $\eta$ in
between is also possible, given the systematic uncertainties in these
abundance determinations.\cite{copi2,kersar} At present, only the
$\Hefour$ abundance inferred from BCGs is reasonably established to be
primordial and even here we can only trust its upper, not lower,
bound, given the empirical linear extrapolation to zero
metallicity. Thus to be {\em conservative} we can only determine the
{\em upper} limits on the parameters $N_\nu$ and $\eta$ corresponding
to the reliable bounds, $Y_{\pr}({\Hefour})<0.25$ (eq.\ref{Ypitl}),
${\Htwo}/{\H}>1.1\times10^{-5}$ (eq.\ref{Dism}) and
${\Liseven}/{\H}<1.5\times10^{-9}$ (eq.\ref{Li7popI}), taking into
account uncertainties in the nuclear cross-sections and the neutron
lifetime by Monte Carlo methods. This exercise finds \cite{kersar}
\begin{equation}
\label{Nnu4.53}
 N_{\nu}^{\max} = 3.75 + 78\ (Y_{\pr}^{\max} - 0.240) ,
\end{equation}
i.e. upto 1.5 additional neutrino species are allowed for $\eta$ at its
lowest allowed value. Conversely, for $N_\nu=3$,
\begin{equation}
 \eta^{\rm max} = [3.28 + 216.4\,(Y_{\pr}^{\max} - 0.240) + 
               34521\,(Y_{\pr}^{\max} - 0.240)^2] \times 10^{-10} ,
\label{etamaxI}
\end{equation}
so that $\eta=8.9\times10^{-10}\Rightarrow\Omega_{\N}=0.033h^{-2}$ is
the maximum allowed value.

The ``crisis'' in BBN identified by Hata \etal~\cite{hata} essentially
arose because they used a chemical evolution model, normalized to
Solar system abundances and convolved with BBN predictions, to infer
that the {\em primordial} abundances were,
${\Htwo}/{\H}=3.5^{+2.7}_{-1.8}\times10^{-5}$ and
${\Hethree}/{\H}=1.2\pm0.3\times10^{-5}$ at $\95cl$. This picks out a
high value of $\eta\approx4.4\times10^{-10}$ which would imply, for
$N_\nu=3$, a higher abundance of $\Hefour$ than their adopted value
(\ref{Ypos}). These authors found that concordance requires
$N_\nu=2.1\pm0.3$ and suggested various exotic possibilities to
achieve this. They also noted that the crisis evaporates if the
systematic uncertainty in the estimate (\ref{Ypos}) of
$Y_{\pr}({\Hefour})$ has been underestimated or if the extent to which
$\Hethree$ survives stellar processing has been overestimated. As we
have seen there are observational grounds for both
suppositions. Overall it is clear that abundances derived from
chemical evolution arguments are suspect and we should only consider
direct observational limits. This is now accepted even by authors who
previously used such arguments to derive bounds such as the one in
eq.(\ref{He3plusD}) on ${\Htwo}+{\Hethree}$, which implies a lower
limit of $\eta\gtrsim2.5\times10^{-10}$ leading to the much advertised
\cite{walker,copi1} constraint $N_\nu\lesssim3.4$.\footnote{When
correlations between the different elemental yields are taken into
account, the actual limit coorresponding to these bounds is even more
stringent,\cite{kraker} $N_{\nu}<3.04$!} Some of them have indeed
conceded \cite{copi2,revisi} that such arguments were not reliable.

\begin{figure}[t]
\epsfxsize\hsize\epsffile{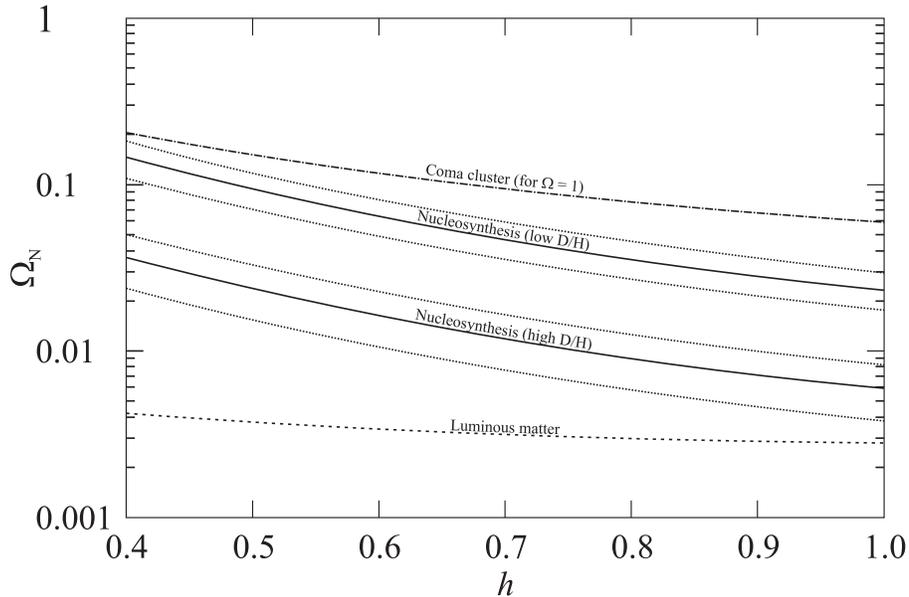}
\caption{The cosmological density parameter in nucleons as a function
  of the Hubble parameter.$^{17}$ The full lines (with dotted
  `$2\sigma$' error bands) show the standard BBN values according as
  whether the primordial $\Htwo$ abundance is taken to be the high
  value $^{44}$ or the low value $^{50}$ measured in QAS. The
  dashed line is the lower limit from an audit $^{3}$ of luminous
  matter in the universe. The dot-dashed line indicates the value
  deduced $^{11}$ from the observed luminous matter in the {\it Coma}
  cluster for $\Omega=1$; it should be lowered by a factor of
  $\Omega^{-1}$ for $\Omega<1$.}
\label{omegan}
\end{figure}

What then are the implications for the nature of the dark
matter?\cite{myrev} For many years it has been stated
\cite{yang,walker,smith,copi1} on the basis of the bound
(\ref{He3plusD}) on ${\Htwo}+{\Hethree}$ combined with the upper bound
from (\ref{Ypos}) on $\Hefour$ that BBN determines the nucleon density
to be $\Omega_{\N}\approx0.011\pm0.0015h^{-2}$. It was argued that
this is significantly higher than the value obtained from direct
observations of luminous matter, suggesting that most nucleons are
dark and, in particular, that much of the dark matter in galactic
halos which contribute \cite{book} $\Omega\approx0.05h^{-1}$ may be
nucleonic. However if the indications of a high primordial $\Htwo$
abundance (eq.\ref{Dlya1}) are correct then as shown in
figure~\ref{omegan} the implied lower value of
$\Omega_{\N}\simeq0.0059\pm0.0011\,h^{-2}$ is closer to its
observational lower limit (for high values of $h$), leaving little
room for the halo dark matter to be nucleonic. Conversely, if the
primordial $\Htwo$ abundance is actually low (eq.\ref{Dlya2}) then the
corresponding value of $\Omega_{\N}=0.023\pm0.0032\,h^{-2}$ would
suggest that the opposite is the case. The results from gravitational
microlensing searches \cite{macho} allow both possibilities at present
and are unlikely to provide a definitive resolution.

More interesting is the comparison with clusters of galaxies which are
observed to have a large nucleonic fraction (eq.\ref{fn}). If
$\Omega=1$, then as shown in figure~\ref{omegan}, the {\sl Coma}
observations can be consistent with standard BBN only for a low
deuterium abundance (and low values of $h$). In fact observations of
large-scale structure and CMB anisotropy do favour high $\Omega_{\N}$
and low $h$ for a critical density universe dominated by cold dark
matter.\cite{doppler} Conversely if the deuterium abundance is
actually high, then to achieve consistency would require
$\Omega\approx0.2$, which is, admittedly, consistent with all dynamic
measurements.\cite{book} The dark matter in {\sl Coma} and other
clusters would then be comparable to that in the individual galactic
halos. However this is not yet a firm argument for $\Omega<1$ since
gravitational lensing observations suggest that the assumption of
hydrostatic equilibrium underestimates the total cluster
mass;\cite{clustermass} for example there may be sources of
non-thermal pressure such as magnetic fields and cosmic rays which
would lower the inferred thermal pressure of the X-ray emitting
plasma.\cite{nontherm} Note that in either case, most of the matter in
the universe {\em must} be non-nucleonic, although not neccessarily
present in our Galactic halo where it can be searched for by direct
experimental means.

The present situation is confusing but it has focussed attention on
systematic errors in measurements of elemental abundances and made it
evident that chemical evolution models are uncertain and results based
on them are not to be trusted. Within a decade $\Omega_{\N}$ is
expected to be known independently to within a few per cent through
measurement of the height of the Doppler peak in the CMB angular power
spectrum.\cite{cmb} Nevertheless precise measurement of light element
abundances is still crucial because primordial nucleosynthesis
provides an unique probe of physical conditions, for example the
particle content, in the early universe. The challenge for observers
is to be ready by then to perform an unprecedented consistency check
of the Standard Models of cosmology and particle physics, and perhaps
to even glimpse what lies beyond.

\section*{Acknowledgments}
It is a pleasure to thank Professor Klapdor-Kleingrothaus and Yorck
Ramachers for the invitation to this enjoyable meeting and Keith Olive
for a stimulating debate. I thank Peter Kernan for an enjoyable
collaboration.

\end{document}